\documentclass[a4paper,11pt]{article}
\usepackage{jcappub} 
\usepackage{booktabs}
\usepackage{tabularx}
\usepackage{threeparttable}
\newcolumntype{L}{>{\raggedright\arraybackslash}X}

\arxivnumber{2512.04604} 
\title{\boldmath Charged Regular Black Holes From Quasi-topological Gravities in $D\ge 5$}

\author[a, b]{Chen-Hao Hao,}
\author[a, b]{Jiliang Jing,}
\author[a, b]{Jieci Wang\note{Corresponding author.}} 

\affiliation[a]{Department of Physics, Key Laboratory of Low Dimensional Quantum Structures and Quantum Control of Ministry of Education, and Synergetic Innovation Center for Quantum Effects and Applications, Hunan Normal University,\\ Changsha, Hunan 410081, P. R. China}
\affiliation[b]{Institute of Interdisciplinary Studies, Hunan Normal University,\\ Changsha, Hunan 410081, P. R. China}

\emailAdd{jcwang@hunnu.edu.cn}

\abstract{The investigation of gravity in higher-dimensional spacetime  has transitioned from a mathematical curiosity to a fundamental framework in theoretical physics, catalyzed by the dimensional requirements of String theory and M-theory.	In this paper, we explicitly construct a spherically symmetric charged black hole solution in $D \ge 5$ dimensions within a gravity theory featuring an infinite tower of higher-curvature corrections. For a given mass and electric charge, the model admits a unique static spherically symmetric solution. We demonstrate that, with an appropriate choice of coupling coefficients $\alpha_n$, the central singularity is progressively mitigated as the correction order increases, ultimately resolving into a globally regular spacetime in the limit of infinite-order corrections. Furthermore, the criteria for the existence of extremal black holes are determined.}

\keywords{Regular black holes, Quasi-topological gravity, Higher dimensions}

\begin{document}
	\maketitle
	\flushbottom
	\section{Introduction}
	
	General Relativity predicts that gravitational collapse leads to the formation of black holes containing physically unacceptable singularities, which indicate a breakdown of causality \cite{Hawking:1970zqf,Penrose:1964wq,Senovilla:1998oua,Penrose:1969pc,Oppenheimer:1939ue}. Since a mature theory of quantum gravity remains elusive, resolving this pathology within classical frameworks becomes crucial. While approaches such as ad hoc modified metric regularizations \cite{Duan:1954bms,Sakharov:1966aja,Bardeen1,Dymnikova:1992ux,Capozziello:2025wwl} or the introduction of exotic matter \cite{Hayward:2005gi,Ayon-Beato:1998hmi,Ayon-Beato:2000mjt,Bronnikov:2000vy,Bronnikov:2000yz,Dymnikova:2004zc,Fan:2016hvf,Lan:2023cvz,Bronnikov:2012ch,Gonzalez:2008wd,Dialektopoulos:2025mfz,Hao:2023igi} have been proposed, they often suffer from a lack of physical motivation or solution instabilities. A more plausible scenario is the modified gravity theories \cite{Berej:2006cc,Junior:2023ixh,Aros:2019quj}. This aligns with the idea that  General Relativity is not a complete theory and that it must be modified in large curvature regimes. Notably, in $D \geq 5$ dimensions, supplementing the Einstein-Hilbert action with infinite-order higher-curvature terms—specifically within the ``quasi-topological gravity'' class \cite{Myers:2010ru,Dehghani:2011vu,Cisterna:2017umf}—has been shown to successfully eliminate the Schwarzschild singularity, offering a promising and general basis for effective gravitational actions \cite{Bueno:2024dgm,Bueno:2024zsx,Bueno:2024eig,Frolov:2024hhe,Hu:2023juh}.
	
	Quasi-topological gravity, as a class of metric theories, investigates effective gravitational actions in spacetime dimensions higher than four. It not only expands the boundaries of existing higher-dimensional gravity models \cite{Tangherlini:1963bw,Myers:1986un}, but also, to a certain extent, aligns with developments in string theory and M-theory \cite{Witten:1995ex,Maldacena:1997re,Witten:1998qj}. Furthermore,  general gravity theories in higher dimensions exhibit a landscape far richer than its four-dimensional counterpart. Unlike the strict uniqueness theorems that constrain four-dimensional black holes to the Kerr family, the higher-dimensional phase space admits a diverse "zoo" of solutions with non-spherical event horizon topologies—such as black rings and black Saturns—and exhibit novel dynamical instabilities like the Gregory-Laflamme instability \cite{Emparan:2001wn,Elvang:2007rd,Gregory:1993vy}. Consequently, investigating gravity solutions in $D \ge 5$ provides the foundational basis for probing these complex gravitational structures and testing the universality of modified gravity theories \cite{Lovelock:1971yv,Boulware:1985wk}.
	
	So far, given that regular spherically symmetric black holes have been established in quasi-topological gravity \cite{Bueno:2024dgm,Bueno:2024zsx}, extending these to charged solutions---analogous to the Schwarzschild-to-Reissner-Nordstr\"om generalization---is a natural progression. In this work, we demonstrate that even with a minimally coupled classical Maxwell field, the gravitational field equations reduce to an algebraic form, yielding general charged spherically symmetric solutions. By imposing mild constraints on the couplings $\alpha_n$, we obtain regular charged black holes under infinite-order corrections. Notably, the higher-derivative corrections effectively absorb the divergence of the linear Maxwell field, manifesting a generic regularization mechanism via the non-perturbative response of the gravitational sector. It is worth noting that \cite{Cano:2020qhy,Cano:2020ezi,Hennigar:2025yqm} first obtained the corresponding charged regular black hole solutions, and \cite{Hennigar:2025yqm} systematically presented the results for both linear and nonlinear electromagnetic fields minimally coupled to this gravitational framework. Our  approach not only simplifies the derivation by reducing the field equations to an algebraic system but also provides a more transparent mechanism for understanding how infinite-order corrections resolve the singularity. Despite the astrophysical tendency for rapid discharge \cite{Gibbons:1975kk}, studying these solutions remains crucial, serving as a valuable analogy for the internal structure of rotating black holes \cite{Poisson:1990eh} and offering insights into string theory and supergravity via the extremal limit \cite{Poisson:1990eh,Strominger:1996sh}.
	
	\section{Model}
	
	The correction terms at each order in the effective gravitational action can be modified through perturbative field redefinitions of the metric, thereby providing the freedom to choose among different bases of curvature invariants.  To facilitate the construction of a simple algebraic system of equations, we follow \cite{Bueno:2024dgm} and select a specific basis of densities defined in $D \ge 5$ dimensions. The corresponding action can be written as
	
	\begin{equation}\label{equ1}
		S = \int \mathrm{d}^D x \sqrt{|g|} \left[ \frac{R}{16\pi G} + \sum_{n=2}^{n_{\text{max}}} \frac{\alpha_n \mathcal{Z}_n}{16\pi G}  + \mathcal{L}_M \right],
	\end{equation} 
	where $g$ is the determinant of the metric tensor $g_{ab}$, $R$ denotes the Ricci scalar, $\mathcal{Z}_n$ represents the Lagrangian density of the n-th order, and $\alpha_n$ is the coupling constant corresponding to the order of correction. The detailed form of $\mathcal{Z}_n$ can be found in \cite{Bueno:2024dgm}. The specific form of the higher-order corrections in Eq.~(\ref{equ1}) is physically motivated by Effective Field Theory (EFT) arguments. It has been conjectured that, all higher-order curvature gravities can be
	characterized as generalized quasi
	topological gravity theories via field redefinitions \cite{Bueno:2019ltp}. This suggests that our model captures the universal features of quantum gravity corrections. However, we acknowledge that a rigorous mathematical proof of this mapping for infinite-order expansions remains elusive. Consequently, the specific infinite series of higher curvature terms chosen in this work should be regarded as a phenomenological ansatz that renders the field equations analytically solvable, rather than an exact derivation from a specific string compactification. The Lagrangian density for the Maxwell field, denoted by
	\begin{equation}
		\label{LM}
		\mathcal{L}_M =-\frac{1}{4}F_{\mu\nu}F^{\mu\nu}.
	\end{equation}
	
	Next, we introduce the static spherically symmetric spacetime metric and the ansatz for the Maxwell field,
	\begin{equation}\label{equ3}
		\mathrm ds^2=-N(r)^2f(r)\mathrm dt^2+\frac{\mathrm dr^2}{f(r)}+r^2\mathrm d\Omega_{D-2}^2,
	\end{equation}
	where $N(r)$ and $f(r)$ are two undetermined functions depend only on the radial distance $r$, and the ansatz for the Maxwell field is given by
	\begin{equation}\label{equ5}
		A_\mu dx^\mu= \Phi(r) dt.
	\end{equation}
	
	After that, we use reduced Lagrangian methods \cite{Fels:2001rv,Deser:2003up} to give the reduced actions for the geometric sector (in terms of $N$ and $\sigma$) and the Maxwell field sector ($\Phi$) 
	\begin{equation}
		I_{grav} = \frac{\Omega_{D-2}}{16\pi G} \int dt dr N(r) \left[r^{D-1}h(\psi)\right]',
	\end{equation}
	\begin{equation}
		I_{em} = \frac{\Omega_{D-2}}{2} \int dt dr \frac{r^{D-2}(\Phi')^2}{N(r)},
	\end{equation}
	where $\psi = \frac{1-f(r)}{r^2}$, $h(\psi) \equiv \psi + \sum_{n=2}^{n_{\text{max}}} \alpha_n \psi^n$, $\alpha_n$ is the he Quasi-topological couplings and the prime $'$ denotes $\frac{d}{dr}$. The function $h(\psi)$ acts as the characteristic polynomial for the theory within the spherically symmetric sector. It compactly encodes the information of the infinite tower of higher-order Lagrangian terms and facilitates the reduction of the complex fourth-order field equations into a single algebraic master equation.
	\begin{itemize}
		\item Variation with respect to $f(r)$:
	\end{itemize}
	
	Note that the Maxwell term $I_{em}$ does not explicitly depend on $f(r)$. Consequently, the result of the variation is identical to that of the vacuum case
	\begin{equation}
		\frac{\delta I_{total}}{\delta f} = \frac{\delta I_{grav}}{\delta f} = 0 \implies h'(\psi)N'(r) = 0.
	\end{equation}
	This implies that we must have 
	$
	N(r) = \text{const} = 1$, which  demonstrates that even with the inclusion of electric charge, the reciprocal relationship between the metric functions $g_{tt}$ and $g_{rr}^{-1}$ is preserved.
	
	\begin{itemize}
		\item Variation with respect to $\Phi(r)$ :
	\end{itemize}
	
	Solving Maxwell's equations by varying with respect to the electric potential $\Phi(r)$ ($\delta I / \delta \Phi = 0$)
	\begin{equation}
		\partial_r \left( \frac{r^{D-2}\Phi'(r)}{N(r)} \right) = 0 \implies \frac{r^{D-2}\Phi'(r)}{N(r)} = -\tilde{Q}.
	\end{equation}
	
	The integration constant $\tilde{Q}$ appearing here is corresponding to the physical electric charge defined according to Gauss's law. Thus, we obtain the electric field strength
	\begin{equation}
		\Phi' = - \frac{N(r) \tilde{Q}}{r^{D-2}} = - \frac{ \tilde{Q}}{r^{D-2}},
	\end{equation}
	we define the physical charge $Q$ as the electric flux at infinity
	\begin{equation}
		Q \equiv \oint_{S^{D-2}} {*F} = \Omega_{D-2} \lim_{r \to \infty} \left( -r^{D-2} \Phi'(r) \right), 
	\end{equation}
	this yields the relationship between the constant and the charge
	\begin{equation}
		\tilde{Q} = \frac{Q}{\Omega_{D-2}}.
	\end{equation}
	
	\begin{itemize}
		\item Variation with respect to $N$. 
	\end{itemize}
	
	Varying the action with respect to the metric function $N(r)$ ($\delta I/\delta N = 0$) yields the master equation. Substituting $\Phi'$ back into the variation of $I_{em}$
	\begin{equation}
		\frac{\delta I_{grav}}{\delta N} + \frac{\delta I_{em}}{\delta N} = 0.
	\end{equation}
	Then we obtain 
	\begin{equation}
		\frac{\Omega_{D-2}}{16\pi G} [r^{D-1}h(\psi)]' = \frac{{Q}^2}{2\Omega_{D-2}r^{D-2}}.
	\end{equation}
	
	Integrating the differential equation obtained, we obtain the expression for the polynomial $h(\psi)$
	\begin{equation}
		\label{ee}
		h(\psi) = \frac{m}{r^{D-1}} - \frac{q}{r^{2D-4}}.
	\end{equation}
	In this equation the integration constant $m$ and the charge parameter $q$ are related to the ADM mass $M$ and the physical charge $Q$ respectively, defined as 
	\begin{equation}
		m = \frac{16\pi G M}{(D-2)\Omega_{D-2}},
	\end{equation}
	\begin{equation}
		q = \frac{8\pi G Q^2}{(D-2)(D-3)\Omega_{D-2}^2}.
	\end{equation}
	
	If our theory is truncated at the $N$-th order ($n_{max} = N$), then $h(\psi)$ is dominated by $\psi^N$ as $\psi \to \infty$
	\begin{equation}
		h(\psi) \approx \alpha_N \psi^N.
	\end{equation}
	At small $r$, the charge term dominates the right-hand side of the algebraic equation (the mass term is subleading), and one can obtain 
	\begin{equation}
		\alpha_N \psi^N \approx - \frac{q}{r^{2D-4}}.
	\end{equation}
	Solving for $\psi$ (and recalling $\psi = \frac{1-f(r)}{r^2}$), we can obtain the asymptotic behavior of the metric function $f(r)$ near the origin   
	\begin{equation}\label{ex}
		f(r) \approx 1 - \left( - \frac{q}{\alpha_N} \right)^{1/N} r^{2 - \frac{2D-4}{N}},
	\end{equation}
	for finite correction orders $N$, the solution exhibits a divergence at the origin $r=0$. However, it is observed that this divergence is progressively mitigated as the order $N$ increases, ultimately achieving global regularity in the limit of infinite order with finite Kretschman scalar $K=R^{\mu\nu\rho\sigma}R_{\mu\nu\rho\sigma}$. In the charged case, we find that the regularity conditions exhibit subtle but crucial differences from those of the pure vacuum solution and the regular center is an anti-de Sitter core. It is instructive to verify the independence of the regular core from the electric charge $q$ by directly examining the asymptotic solution Eq.~(\ref{ex}). Taking the infinite-order limit $N \to \infty$, the exponent of the radial coordinate in the metric function $f(r)$ trivially approaches $2$. Regarding the coefficient, for any non-vanishing finite charge $q$, the term $|q|^{1/N}$ converges to unity, indicating that the specific intensity of the source becomes irrelevant in this limit. On the other hand, to ensure the geometry remains regular at the center, the coupling coefficients $\alpha_N$ must be such that the term $|\alpha_N|^{1/N}$ converges to a finite non-zero constant (characterizing the convergence radius of the series). Consequently, the effective cosmological constant of the core is determined solely by the asymptotic behavior of the gravitational couplings, independent of the electric charge $q$. The mechanism underlying this transition of the regular core can be found in \cite{Hennigar:2025yqm}.
	
	Unlike the vacuum case \cite{Bueno:2024dgm}, the source term for a charged black hole covers the entire real axis. Therefore, the characteristic function $h(\psi)$ must be able to cover the full dynamic range of the source term. To construct an $h(\psi)$ that satisfies the aforementioned range requirements, the sequence of coupling constants $\alpha_n$ must satisfy the following specific algebraic properties: To allow $h(\psi)$ to take negative values and approach $-\infty$, the series must be dominated by odd powers. The most natural implementation is to require $h(\psi)$ to be an odd function:
	\begin{equation}\label{20}
		\alpha_{2k} = 0 \quad (\forall k \in \mathbb{N}),
	\end{equation}
	to ensure the monotonicity of $h(\psi)$ within its domain (monotonicity not only guarantees the uniqueness of the solution but also prevents the emergence of ``branch singularities'' that could arise from multiple solution branches), the remaining odd-order coupling constants must be non-negative
	\begin{equation}\label{21}
		\alpha_{2k+1} \ge 0 \quad (\forall k \in \mathbb{N}),
	\end{equation}
	to achieve metric regularity at $r=0$ (i.e., $f(r) \approx 1 + \text{C} \cdot r^2$), $\psi$ must approach a finite constant when the source term diverges. This requires the power series of $h(\psi)$ to have a finite radius of convergence and to diverge at this radius
	\begin{equation}\label{22}
		\lim_{k\to\infty} (\alpha_{2k+1})^{\frac{1}{2k+1}} = C > 0.
	\end{equation}
	
	A very natural choice can be found in  \cite{Bueno:2024dgm}: $\frac{(1 - (-1)^n)\Gamma \left(\frac{n}{2}\right)}{2\sqrt{\pi}\Gamma \left(\frac{n+1}{2}\right)} \alpha^{n-1} = \alpha_n$. It is important to note that the specific sequence of coupling constants $\alpha_n$ is primarily for mathematical convenience to derive a simple closed-form master equation. The regularization mechanism does not rely on fine-tuning these specific values. The generic requirement is that the characteristic function $h(\psi)$ must have a finite radius of convergence $\psi_c$ and diverge at this radius. This ensures that the field $\psi$ saturates to a finite value as the source term diverges. Within a broad class of coupling sequences that satisfy the asymptotic conditions Eq.(\ref{20}) - Eq.(\ref{22}), this saturation—and the resulting regular core—persists, although the detailed profile of the solution depends on the specific choice of coefficients. Substituting the $\alpha_n$ into Eq.~(\ref{ee}), we obtain the following equations to be solved for different orders of correction

	\noindent 1. $n=1$:
	\begin{subequations}
		\label{eq:n_equals_1}
		\begin{align}
			&\psi = \frac{m}{r^{D-1}} - \frac{q}{r^{2D-4}}, \\
			&f(r) = 1 - \frac{m}{r^{D-3}} + \frac{q}{r^{2D-6}}, \\
			&f(r)\big|_{r \to 0} \approx 1 + \frac{q}{r^{2D-6}}, \\
			&K\big|_{r \to 0} \sim \psi^2 \approx \frac{q^2}{r^{4D-8}}.
		\end{align}
	\end{subequations}
	
	\noindent 2. $n=3$:
	\begin{subequations}
		\label{eq:n_equals_3}
		\begin{align}
			&\psi + \frac{1}{2}\alpha^2 \psi^3 = \frac{m}{r^{D-1}} - \frac{q}{r^{2D-4}}, \\
			&f(r) = 1 - \frac{r^2}{\sqrt{3}\alpha} \left[ (\mathcal{L} + \mathcal{K})^{1/3} + (\mathcal{L} - \mathcal{K})^{1/3} \right], \\
			&\mathcal{L} = \frac{3\sqrt{3}\alpha}{2} \left( \frac{m}{r^{D-1}} - \frac{q}{r^{2D-4}} \right), \nonumber \\
			&\mathcal{K} = \sqrt{1 + \frac{27\alpha^2}{4} \left( \frac{m}{r^{D-1}} - \frac{q}{r^{2D-4}} \right)^2}, \\
			&f(r)\big|_{r \to 0} = 1 - r^2\psi \approx 1 + \left(\frac{2q}{\alpha^2}\right)^{1/3} r^{\frac{10-2D}{3}}, \\
			&K\big|_{r \to 0} \approx r^{-\frac{4D-8}{3}}.
		\end{align}
	\end{subequations}
	
	\noindent 3. $n=5$:
	\begin{subequations}
		\label{eq:n_equals_5}
		\begin{align}
			&\psi + \frac{1}{2}\alpha^2 \psi^3 + \frac{3}{8}\alpha^4 \psi^5 = \frac{m}{r^{D-1}} - \frac{q}{r^{2D-4}}, \\
			&f(r)\big|_{r \to 0} \approx 1 + \mathcal{C}_5 \cdot r^{\frac{14-2D}{5}}, \\
			&K\big|_{r \to 0} \approx r^{-\frac{4D-8}{5}}.
		\end{align}
	\end{subequations}
	
	\noindent 4. $n=\infty$:
	\begin{subequations}
		\label{eq:n_infinity}
		\begin{align}
			&\frac{\psi}{\sqrt{1 - \alpha^2 \psi^2}} = \frac{m}{r^{D-1}} - \frac{q}{r^{2D-4}}, \\
			&f(r)= 1 - r^2 \left[ \frac{mr^{D-3} - q}{\sqrt{r^{2(2D-4)} + \alpha^2 (mr^{D-3} - q)^2}} \right], \\
			&f(r)\big|_{r \to 0} \approx 1 + \frac{r^2}{\alpha}, \\
			&K\big|_{r=0} = \frac{2D(D-1)}{\alpha^2}.
		\end{align}
	\end{subequations}
	
	For $n=1$, the solution recovers the standard Tangherlini-Reissner-Nordström black hole in $D \ge 5$ Einstein gravity \cite{Wiltshire:1985us}. Explicit analytic solutions can be derived for both the $n=3$ case and the infinite-order limit; however, for $n \ge 5$, no explicit closed-form expressions can be found. 
	
	\section{Solution in D = 5}
	
\begin{figure}[htbp]
		\centering
		\includegraphics[width=0.7\linewidth]{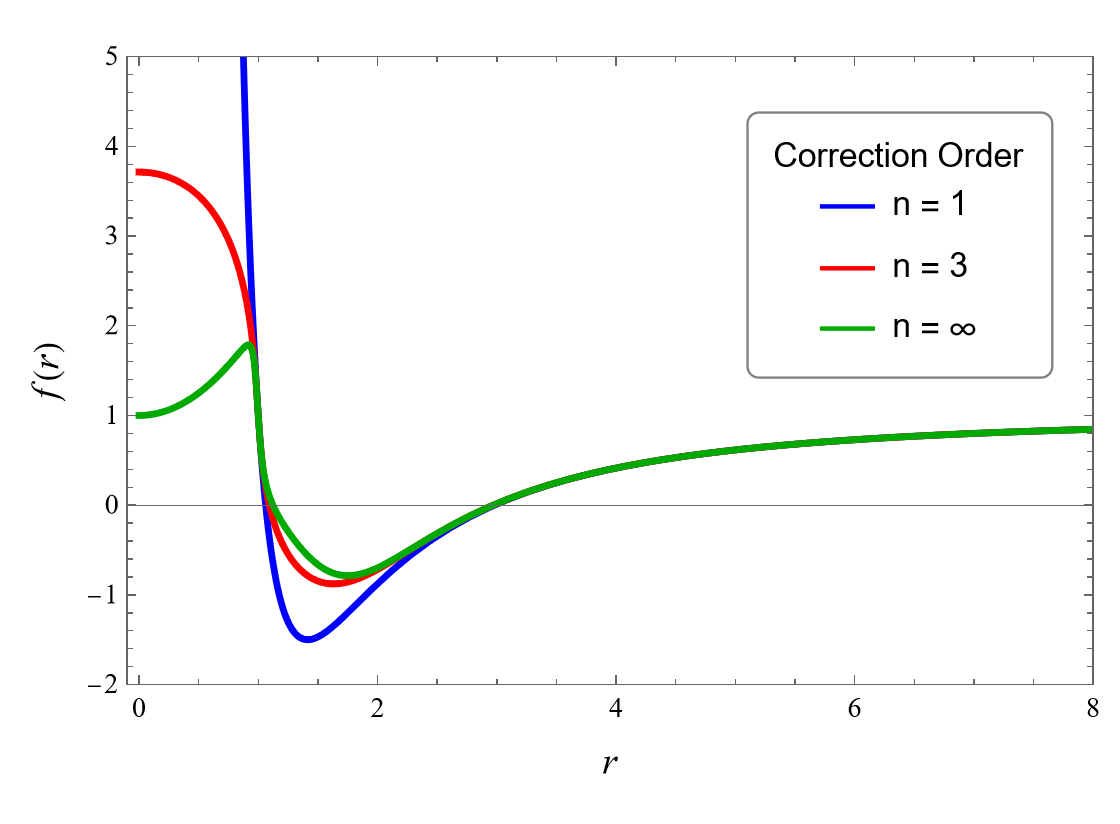}
		\caption{The profiles of the metric function $f(r)$ for different correction orders $n = 1,3,\infty$, with the parameters fixed at $m=10$, $q=10$, and $\alpha=1$. All physical quantities are expressed in Planck units ($G=c=1$). Length quantities are normalized by the coupling scale $\sqrt{\alpha}$, making the mass $m$ and charge $q$ dimensionless parameters.}
		\label{p1}
	\end{figure}

	It is instructive to analyze the obtained solutions within a specific spacetime dimension. In the subsequent discussion, we fix the spacetime dimension to $D=5$. The explicit forms and properties of the solutions in this case are summarized in Table~\ref{tab:1}. Furthermore, in Fig.~\ref{p1}, we present the profiles of the metric function $f(r)$ for different correction orders, with the parameters fixed at $m=10$, $q=10$, and $\alpha=1$.
	
    \begin{table}[htbp]
    \centering
    \begin{threeparttable} 
    \caption{The explicit forms of $f(r)$ and the corresponding asymptotic behavior of $f(r)$ and $K$ at $r = 0$.}
    \label{tab1}
    \small
    
    \newcolumntype{L}{>{\raggedright\arraybackslash}X}
    \newcolumntype{C}{>{\centering\arraybackslash}c}

    \begin{tabularx}{\textwidth}{c L C C} 
        \toprule
        \textbf{Correction Order} & \textbf{Metric Function} $f(r)$ & \textbf{Expansion of} $f(r)$ & \textbf{Expansion of} $K$ \\ 
        \midrule
        
        $n=1$ & 
        $f(r) = 1 - \frac{m}{r^2} + \frac{q}{r^4}$ & 
        $\sim 1 + \frac{q}{r^4}$ & 
        $\sim r^{-12}$ \\ 
        \addlinespace[1.5ex]
        
        $n=3$ & 
        $\begin{aligned}
            f(r) &= 1 - \frac{1}{\sqrt{3}\alpha} \Bigl[ \\
            &\quad (F_1)^{1/3} - (F_2)^{1/3} \Bigr]
        \end{aligned}$ & 
        $\sim 1 + (\frac{2q}{\alpha^2})^{1/3}$ & 
        $\sim r^{-4}$ \\ 
        \addlinespace[1.5ex]
        
        $n=\infty$ & 
        $f(r) = 1 - \frac{r^2(mr^2-q)}{\sqrt{r^{12}+\alpha^2(mr^2-q)^2}}$ & 
        $\sim 1 + \frac{r^2}{\alpha}$ & 
        Finite ($\frac{40}{\alpha^2}$) \\ 
        
        \bottomrule
        \label{tab:1}
    \end{tabularx}

    \begin{tablenotes}
        \footnotesize
        \item \textbf{Note:} For the $n=3$ case, the auxiliary functions $F_1$ and $F_2$ are defined as:
        \item $F_1 = 3\sqrt{3}\alpha(mr^2-q) + \sqrt{8r^{12} + 27\alpha^2(mr^2-q)^2}$
        \item $F_2 = 3\sqrt{3}\alpha(mr^2-q) - \sqrt{8r^{12} + 27\alpha^2(mr^2-q)^2}$
    \end{tablenotes}

\end{threeparttable} 
\end{table}
	
	\begin{itemize}
		\item Horizon Structure and Critical Mass 
	\end{itemize}
	
	To analyze the causal structure of the regular solution under $n = \infty$, we investigate the existence of event horizons, which are defined as the roots of the metric function $f(r_h)=0$. Based on the regular solution derived for $D=5$, the horizon equation implies
	\begin{equation}
		1 = \frac{r_h^2 (m r_h^2 - q)}{\sqrt{r_h^{12} + \alpha^2 (m r_h^2 - q)^2}}.
	\end{equation}
	This equation can be inverted to express the mass parameter $m$ as a function of the horizon radius $r_h$ and the charge parameter $q$
	\begin{equation}
		\label{eq:mass_function}
		m(r_h) = \frac{q}{r_h^2} + \frac{r_h^4}{\sqrt{r_h^4 - \alpha^2}}.
	\end{equation}
	The existence of real solutions requires $r_h > \alpha^{1/2}$. For a fixed charge $q$, the number of horizons is determined by the local extrema of the function $m(r_h)$. The condition for an extremal black hole, where the inner and outer horizons coincide, is given by $\partial m / \partial r_h = 0$. This leads to a system of parametric equations describing the critical mass $m_{ext}$ and critical charge $q_{ext}$ in terms of the horizon radius parameter $R \equiv r_h^4$
	\begin{equation}
		\label{eq:critical_params}
		\begin{aligned}
			q_{ext}(R) &= \frac{R^{3/2} (R - 2\alpha^2)}{(R - \alpha^2)^{3/2}}, \\
			m_{ext}(R) &= \frac{R (2R - 3\alpha^2)}{(R - \alpha^2)^{3/2}}.
		\end{aligned}
	\end{equation}
	Here, physical positivity of the charge ($q>0$) imposes a constraint on the horizon radius: $R > 2\alpha^2$. These equations define a critical curve in the $m-q$ phase space that separates black holes from regular solitons. Specifically:
	\begin{itemize}
		\item Case I ($m > m_{ext}(q)$): The metric function has two distinct positive roots, corresponding to a regular non-extremal black hole with an event horizon and a Cauchy horizon.
		\item Case II ($m = m_{ext}(q)$): The metric has a double root, representing an extremal regular black hole with zero surface gravity.
		\item Case III ($m < m_{ext}(q)$): No horizons exist ($f(r) > 0$ everywhere). The spacetime represents a charged regular soliton (or a naked regular core), where the entire manifold is accessible to asymptotic observers.
	\end{itemize}
	
	\begin{figure}[htbp]
		\centering
		\includegraphics[width=0.7\linewidth]{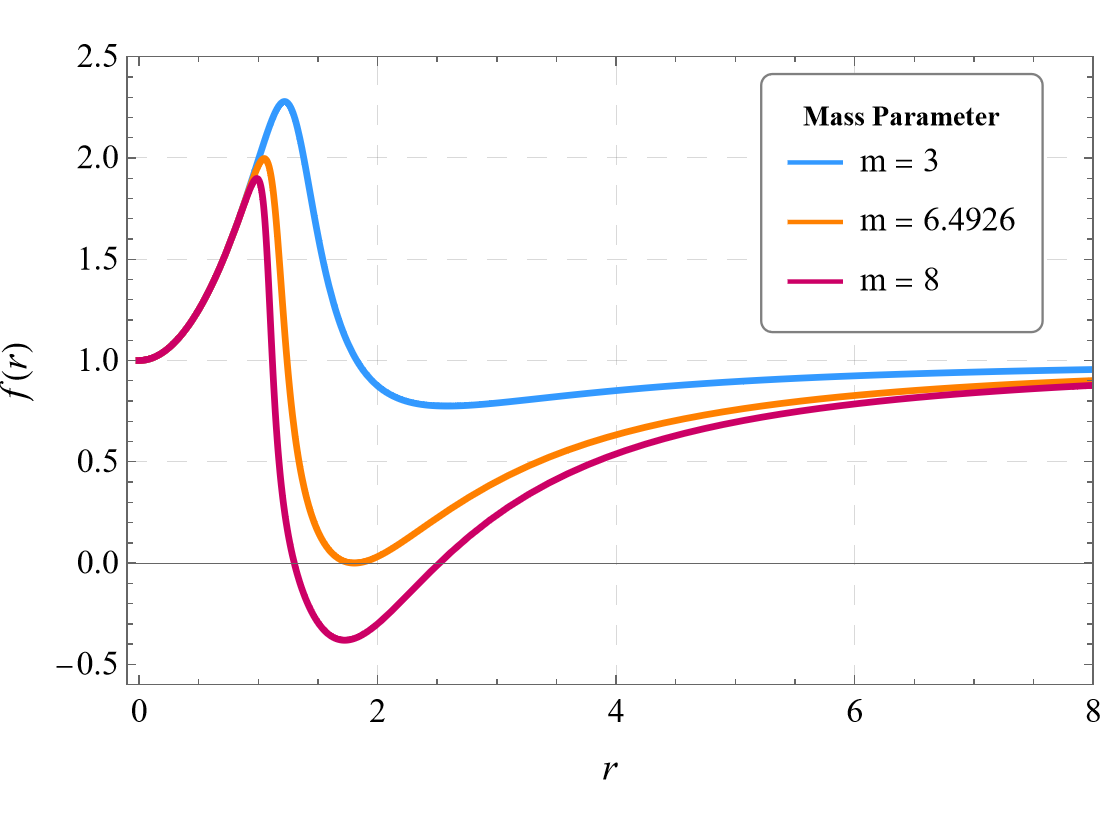}
		\caption{The profiles of the metric function $f(r)$ for $n = \infty$, $q=10$ and $\alpha=1$, with the mass parameter $m=3, 6.4926, 8$. All physical quantities are expressed in Planck units ($G=c=1$). Length quantities are normalized by the coupling scale $\sqrt{\alpha}$, making the mass $m$ and charge $q$ dimensionless parameters.}
		\label{p2}
	\end{figure}
	\begin{figure}[htbp]
		\centering
		\includegraphics[width=0.7\linewidth]{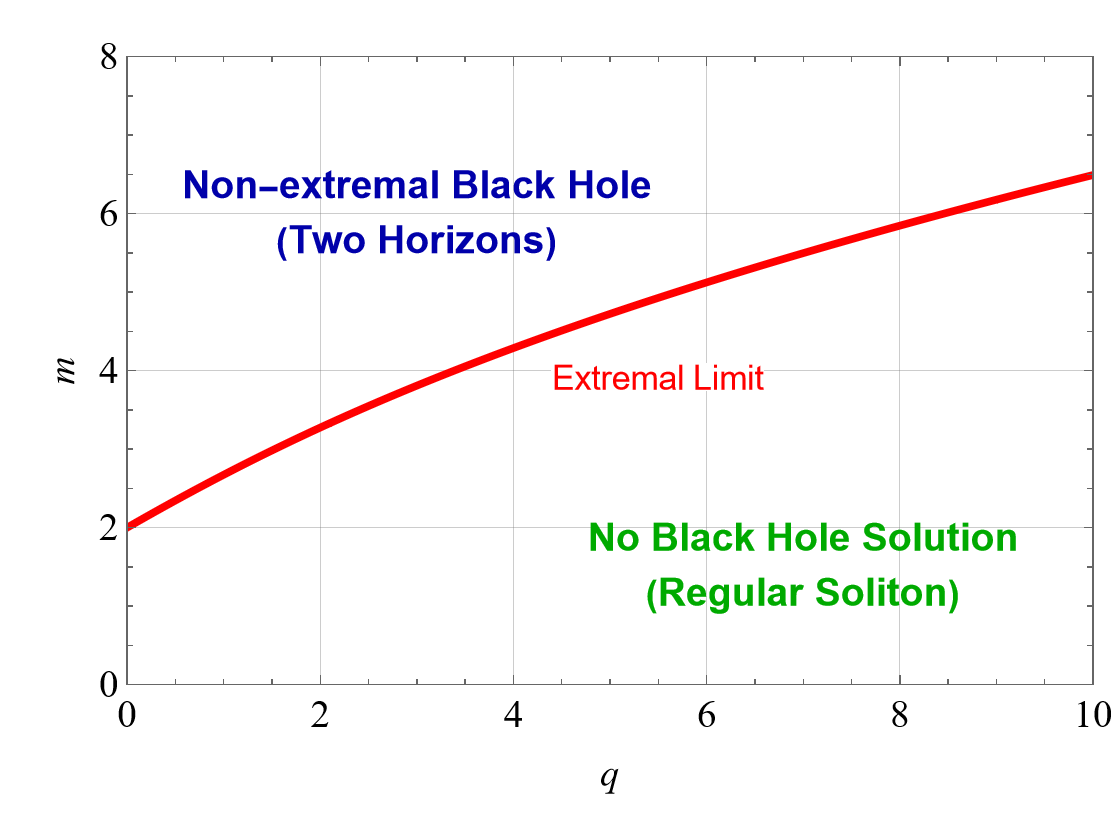}
		\caption{Phase Diagram of 5D Charged Regular Black Holes ($\alpha=1$). All physical quantities are expressed in Planck units ($G=c=1$). Length quantities are normalized by the coupling scale $\sqrt{\alpha}$, making the mass $m$ and charge $q$ dimensionless parameters.}
		\label{p3}
	\end{figure}
	\begin{figure}[htbp]
		\centering
		\includegraphics[width=0.7\linewidth]{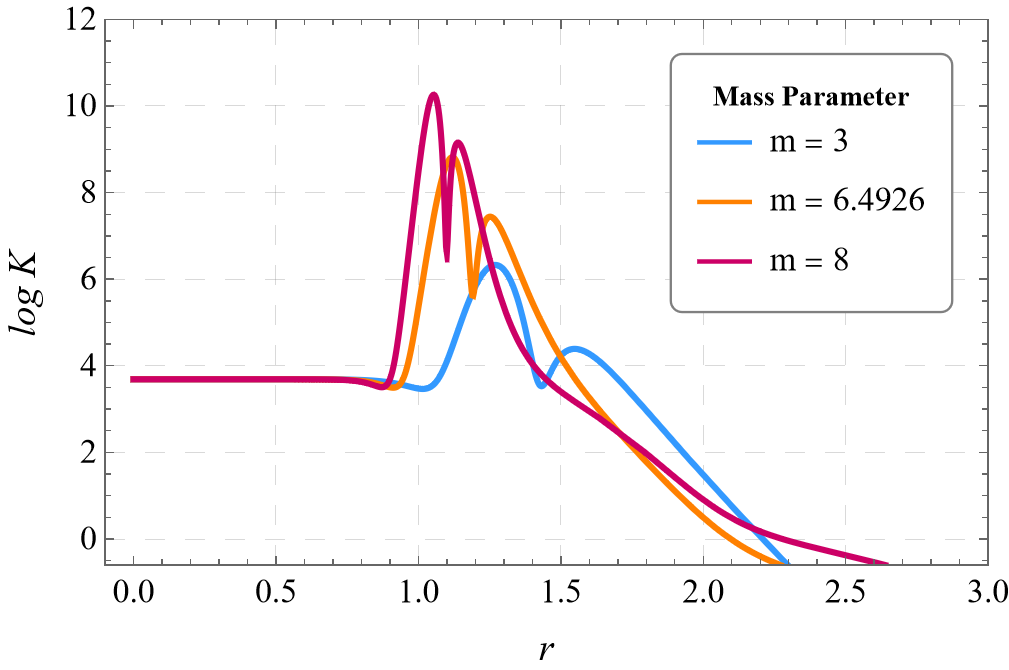}
		\caption{The profiles of the Kretschmann scalar $K$ for correction orders $n =\infty$, with the parameters fixed at $q=10$, $\alpha=1$ and $m = 3, 6.4926, 8$. All physical quantities are expressed in Planck units ($G=c=1$). Length quantities are normalized by the coupling scale $\sqrt{\alpha}$, making the mass $m$ and charge $q$ dimensionless parameters.}
		\label{p4}
	\end{figure}
	It is worth noting that in the limit of vanishing higher-curvature corrections ($\alpha \to 0$), the parametric equations reduce to $q_{ext} \approx R$ and $m_{ext} \approx 2\sqrt{R}$, yielding the relation $m_{ext}^2 = 4q_{ext}$. This recovers the standard extremality condition for a five-dimensional Reissner-Nordstr\"om black hole in General Relativity. Furthermore, it is observed that the formation of a black hole requires the mass parameter $m$ to exceed a critical lower bound. This threshold value increases with the charge $q$, reducing to $m_{min}=2$ in the limit $q \to 0$. In Fig.~\ref{p2}, we present the profiles of the metric function $f(r)$ for fixed parameters $q=10$ and $\alpha=1$ with several representative values of $m$, and Fig.~\ref{p3} presents five-dimensional charged regular black holes with $\alpha=1$. The red solid line represents the extremal limit $m = m_{ext}(q)$, where the inner and outer horizons coincide. The region above the curve (blue) corresponds to non-extremal black holes with two horizons, while the region below (green) represents regular solitons with no event horizon. Fig.~\ref{p4} displays the corresponding Kretschmann scalar. It is important to emphasize that, in contrast to the pure vacuum solution, the Kretschmann scalar for the charged solution exhibits pronounced variations at certain radius, rather than maintaining a globally smooth profile. It is important to clarify that the non-monotonic behavior and sharp peaks observed in the Kretschmann scalar are physical features intrinsic to the regularization mechanism, rather than numerical artifacts. This behavior is correlated with the drastic evolution of the metric function $f(r)$ presented in Fig.\ref{p2}. Specifically, in the transition region where the geometry shifts from the classical exterior to the saturated core, $f(r)$ undergoes a rapid, steep decrease. Since the Kretschmann scalar involves the second derivatives of the metric ($K \sim (f'')^2$), this sharp gradient in $f(r)$ naturally induces a drastic spike in the curvature invariants. Thus, the oscillation in curvature precisely marks the "curvature wall" where the metric function experiences its most significant adjustment to avoid the singularity. Nevertheless, the Kretschmann scalar remains finite throughout the entire spacetime.
	
	\section{Discussion}
	
	In this work, we constructed a novel class of static, spherically symmetric charged black hole solutions in $D \ge 5$ dimensions using quasi-topological gravity with infinite-order curvature corrections. Extending the methods of \cite{Bueno:2024dgm,Cano:2020ezi,Cano:2020qhy,Hennigar:2025yqm}, we derived an algebraic master equation for the metric. Our analysis shows that a physically motivated choice of coupling coefficients $\alpha_n$ resolves the classical Reissner-Nordström singularity. The interplay between charge $q$ and mass $m$ reveals a rich phase structure, comprising regular non-extremal black holes, regular extremal black holes, and horizonless charged solitons.
	
	A key feature of our model is the physical origin of its regularity. Unlike charged regular black hole models such as the Bardeen and Hayward classes, which rely on Non-Linear Electrodynamics (NLED)—to enforce regularity, our construction retains the linear Maxwell field with its divergent energy density. Regularity emerges instead from the non-perturbative response of the gravitational sector: high-order terms saturate the curvature at the scale $\alpha$, absorbing the Maxwell field's divergence and enforcing an anti-de Sitter core. This demonstrates that infinite-order curvature corrections—characteristic of quantum gravity effective field theories—can inherently heal spacetime pathologies without exotic matter.
	
	Future research should prioritize analysis to assess stability, the Quasi-topological Gravity framework ensures that the master equation for the metric is algebraic, thereby avoiding Ostrogradsky ghosts in the static spherically symmetric sector. While a full perturbative stability analysis is beyond the scope of this work, previous studies on similar modified gravities suggest healthy behavior \cite{Konoplya:2024hfg}. Future work will focus on a comprehensive analysis to assess dynamical stability \cite{Nashed:2021pah,Zhao:2023uam}. Additionally, investigating thermodynamic properties will offer insights into their quantum gravitational nature \cite{Cisterna:2025vxk,Wang:2024zlq} and we encourage readers to consult the systematic study presented in \cite{Hennigar:2025yqm}. Finally, extending this construction to rotating solutions remains a challenging but rewarding goal \cite{Burger:2019wkq,Adair:2020vso,Corral:2025yvr,Liu:2025iyl,Bueno:2025qjk,Nashed:2018cth}, as astrophysical black holes possess angular momentum.
	
	\section*{Acknowledgements}
	This work was supported by the National Natural Science Foundation of China under Grants  No. 12475051,  No. 12375051, and No. 12421005; the science and technology innovation Program of Hunan Province under grant No. 2024RC1050; and the Natural Science Foundation of Hunan Province under grant No. 2026JJ20019.

\end{document}